\documentclass[%
 reprint,
 amsmath,amssymb,
 aps,
]{revtex4-2}
\usepackage{subfigure}
\usepackage{graphicx}% Include figure files
\usepackage{dcolumn}% Align table columns on decimal point
\usepackage{bm}% bold math
\usepackage{graphicx,subfigure,float}
\usepackage[T1]{fontenc}
\begin{document}

\preprint{APS/123-QED}
%\title{Characterizing the cold $^{88}$Sr atoms using the $5s5p^{3}$P$_{2}$ - $5s4d^{3}$D$_{1}$ repumping and Land$\acute{e}$ g factor measurement}% Force line breaks with \\

\title{Novel repumping on $^{3}$P$_{0}$$\rightarrow$$^{3}$D$_{1}$ for Sr magneto-optical trap and Land\'{e} $g$ factor measurement of $^{3}$D$_{1}$}% Force line breaks with \\
%in order to verify
%we point out that
%Note that
%We specifically demonstrate that
%depict
%atom cloud
%Rabi broadening
%why measure 3D1 lande g factor? 98.2\% of the dynamic shift is caused by the 3P0 →3D1 transition at 2.603 µm, and uncertainty
%in the Einstein A coefficient for this resonance dominates the uncertainty in νdyn( from 3D1 lifetime measurement Jun Ye's paper)
\author{Shengnan Zhang$^{1}$}
%\affiliation{
 %Midlands Ultracold Atom Research Centre, School of Physics and Astronomy, University of Birmingham, Edgbaston, Birmingham B15 2TT, United Kingdom
%}
% \altaffiliation[Also at ]{}%Lines break automatically or can be forced with \\
\author{Preetam Ramchurn$^{1}$}
%\affiliation{
 %Midlands Ultracold Atom Research Centre, School of Physics and Astronomy, University of Birmingham, Edgbaston, Birmingham B15 2TT, United Kingdom
%}
\author{Marco Menchetti$^{1}$}
%\affiliation{
%Midlands Ultracold Atom Research Centre, School of Physics and Astronomy, University of Birmingham, Edgbaston, Birmingham B15 2TT, United Kingdom
%}
\author{Qasim Ubaid$^{1,2}$}
%\affiliation{
 %Midlands Ultracold Atom Research Centre, School of Physics and Astronomy, University of Birmingham, Edgbaston, Birmingham B15 2TT, United Kingdom
%}

\author{Jonathan Jones$^{1}$}
%\affiliation{
 %Midlands Ultracold Atom Research Centre, School of Physics and Astronomy, University of Birmingham, Edgbaston, Birmingham B15 2TT, United Kingdom
%}
\author{Kai Bongs$^{1}$}%
%\affiliation{
 %Midlands Ultracold Atom Research Centre, School of Physics and Astronomy, University of Birmingham, Edgbaston, Birmingham B15 2TT, United Kingdom
%}
\author{Yeshpal Singh$^{1}$}%
 \email{Y.Singh.1@bham.ac.uk}
\affiliation{
 $^{1}$ Midlands Ultracold Atom Research Centre, School of Physics and Astronomy, University of Birmingham, Edgbaston, Birmingham, B15 2TT, United Kingdom
}
\affiliation{
 $^{2}$ Department of Physics, University of Kerbala, Karbala, 56001, Iraq
}
%\affiliation{$^{2}$ 
%Department of Physics,
%University of Kerbala, Freaha, Kerbala, 1125, Iraq


\begin{abstract}
We realize an experimental facility for cooling and trapping strontium (Sr) atoms and measure the Land\'{e} $g$ factor of $^{3}$D$_{1}$ of $^{88}$Sr. Thanks to a novel repumping scheme with the $^{3}$P$_{2}$$\rightarrow$$^{3}$S$_{1}$ and $^{3}$P$_{0}$$\rightarrow$$^{3}$D$_{1}$ combination and the permanent magnets based self-assembled Zeeman slower, the peak atom number in the continuously repumped blue MOT is enhanced by a factor of 15 with respect to the non-repumping case, and reaches $\sim$1 billion. Furthermore, using the resolved-sideband Zeeman spectroscopy, the Land\'{e} $g$ factor of $^{3}$D$_{1}$ is measured to be 0.4995(88) showing a good agreement with the theoretical value of 0.4988. The results will have an impact on various applications including atom laser, dipolar interactions, quantum information and precision measurements.
\end{abstract}

\maketitle

%\tableofcontents
\section{Introduction}
The growing importance of laser-cooled atomic strontium (Sr) is starting to emerge in diverse fields, where in particular large atom numbers and high density are required, such as precision metrology\cite{Nuovo,RevLud,TargatLod,Katori,Andrews,Pizzoca,Zawadadd,IanMar,YeJunNAT}, continuous Bose-Einstein condensation (BEC)\cite{SBHuangEC,SBEC}, quantum simulation\cite{manybody,quanibokh,GroScien} and prospects towards long range dipolar interactions\cite{MarkSa,longrange,Yusingle}. The barrier to the realization of large atom numbers is considerable atom losses during the first cooling and trapping stage. The involved energy levels of Sr are shown in Fig. 1. When operated on the  $^{1}$S$_{0}$$\rightarrow$$^{1}$P$_{1}$ cooling transition, atoms can decay to $^{3}$P$_{1,2}$ states via $^{1}$D$_{2}$. The $^{3}$P$_{1}$ atoms will decay back to the ground state, and hence closing the cooling cycle, while the $^{3}$P$_{2}$ state is essentially causing atoms to be lost from the MOT due to the 17$\,$min lifetime of $^{3}$P$_{2}$\cite{Dereviankodd}. Thus, we need repumping to tackle this issue. So far various repumping schemes for Sr have been demonstrated including single-laser repumping schemes addressing transitions to higher-lying states, such as $^{1}$D$_{2}$$\rightarrow$$5s6p$$^{1}$P$_{1}$\cite{Kurosu}, $5s5p$$^{3}$P$_{2}$$\rightarrow$$5snd$$^{3}$D$_{2}$,$5p^{2}$$^{3}$P$_{2}$\cite{IanMar,Poli,Specacta,Mickelson,Stellmer,FUYU}, and the most common dual repumping scheme combining $^{3}$P$_{0,2}$$\rightarrow$$^{3}$S$_{1}$ at 679$\,$nm and 707$\,$ nm, respectively\cite{Dinneen,Loftus1}. 
%To the best of our knowledge, so far, however, all achieved atom numbers in MOTs are much lower than 10$^{9}$ below 600$^{\circ}$C of the oven temperature.

In this paper we explore a novel dual repumping scheme for Sr, where the combination of $^{3}$P$_{0}$$\rightarrow$$^{3}$D$_{1}$ at 2.6$\,\mu$m and $^{3}$P$_{2}$$\rightarrow$$^{3}$S$_{1}$ at 707$\,$nm is used. The repumping scheme together with a self-assembled Zeeman slower allows loading 1 billion atoms into a continuously repumped MOT at our oven temperature of 560$^{\circ}$C. Similar atom numbers have been reported in the past\cite{Courti03}. However, these measurements required relatively higher temperature of their oven. In addition, we measure the Land\'{e} $g$ factor of the  $^{3}$D$_{1}$ state using the resolved-sideband Zeeman spectroscopy, and the result shows a good agreement with the theoretically calculated value. These results pave the way towards future applications using these short-IR lasers to explore more physics in Sr, such as long range dipolar interactions\cite{longrange}.  %of $^{3}$P$_{0}$$\rightarrow$$^{3}$D$_{1}$ at 2.6$\,\mu$m combined with $^{3}$P$_{2}$$\rightarrow$$^{3}$S$_{1}$ at 707$\,$nm which has never done before. Calculated from the branching ratios of $^{3}$D$_{1}$$\rightarrow$$^{3}$P$_{0,1,2}$ as shown in Fig. 1, the repumping efficiency of $^{3}$P$_{0}$$\rightarrow$$^{3}$D$_{1}$ is much higher than that of  $^{3}$P$_{0}$$\rightarrow$$^{3}$S$_{1}$. Together with an optimised self-assembled Zeeman slower, this repumping strategy enables to dramatically enhance the atom number in the MOT. The Zeeman slower used in our system is assembled with spherical neodymium iron boride (NdFeB) magnets, resulting in compactness and zero power consumption. In addition, we measure the Land\'{e} $g$ factor of $^{3}$D$_{1}$ using the resolved-sideband Zeeman spectroscopy, and the result shows a good agreement with the theoretically calculated value. These results pave the way towards future applications using these short-IR lasers to explore more physics in Sr, such as long range dipolar interactions\cite{longrange}. 

\begin{figure}[ht]
\includegraphics[scale=0.38]{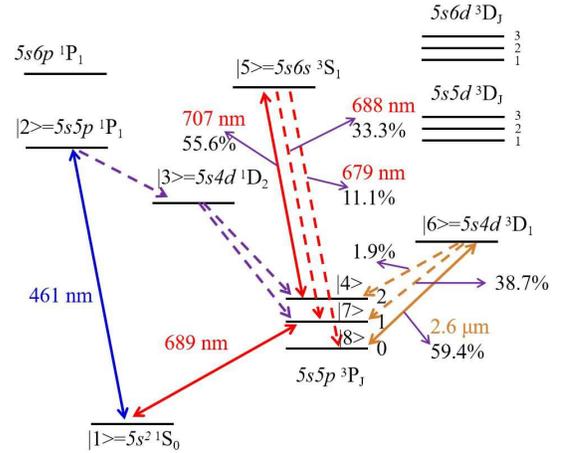}% Here is how to import EPS art
\caption{\label{fig:epsart} A simplified electronic level structure of $^{88}$Sr. The first and second cooling stages work on the $^{1}$S$_{0}$$\rightarrow$$^{1}$P$_{1}$ at 461$\,$nm and $^{1}$S$_{0}$$\rightarrow$$^{3}$P$_{1}$ at 689$\,$nm, respectively. The 707$\,$nm and 679$\,$nm transitions can be used for repumping. In our case we use 707$\,$nm and 2.6$\,\mu$m ($^{3}$P$_{0}$$\rightarrow$$^{3}$D$_{1}$) for repumping. The branching ratios of $^{3}$S$_{1}$$\rightarrow$$^{3}$P$_{0,1,2}$ and $^{3}$D$_{1}$$\rightarrow$$^{3}$P$_{0,1,2}$ are 11.1$\%$, 33.3$\%$, 55.6$\%$, 59.4$\%$, 38.7$\%$ and 1.9$\%$, respectively.}
\end{figure}

\section{Experimental setup}
%Figure 1 shows the simplified electronic level structure of $^{88}$Sr and the transitions relevant for laser cooling are indicated. The cooling and trapping for strontium can be divided into two steps. The first pre-cooling employs the strong $^{1}$S$_{0}$-$^{1}$P$_{1}$ 461 nm transition, of which the saturation intensity is 40 mW/cm$^{2}$. The atoms' temperature at this stage can only be lowered to milli-Kelvin as the consequence of 32 MHz natural linewidth. The Doppler limit temperature is 720 $\mu$K. In this process, the atoms in $^{1}$P$_{1}$ will decay to $^{1}$D$_{2}$ and then $^{3}$P$_{1}$ and $^{3}$P$_{2}$. Almost all the current Sr blue MOT scheme is using 707 nm and 679 nm lasers to repump the $^{3}$P$_{2}$ atoms to $^{3}$P$_{1}$. Whereas, we apply 2.6 $\mu$m laser instead of 679 nm for its higher pumping efficiency. The temperature of the cold atoms after the first stage cooling is still high for the optical lattice experiment. 
\begin{figure*}[ht]
\includegraphics[scale=0.7]{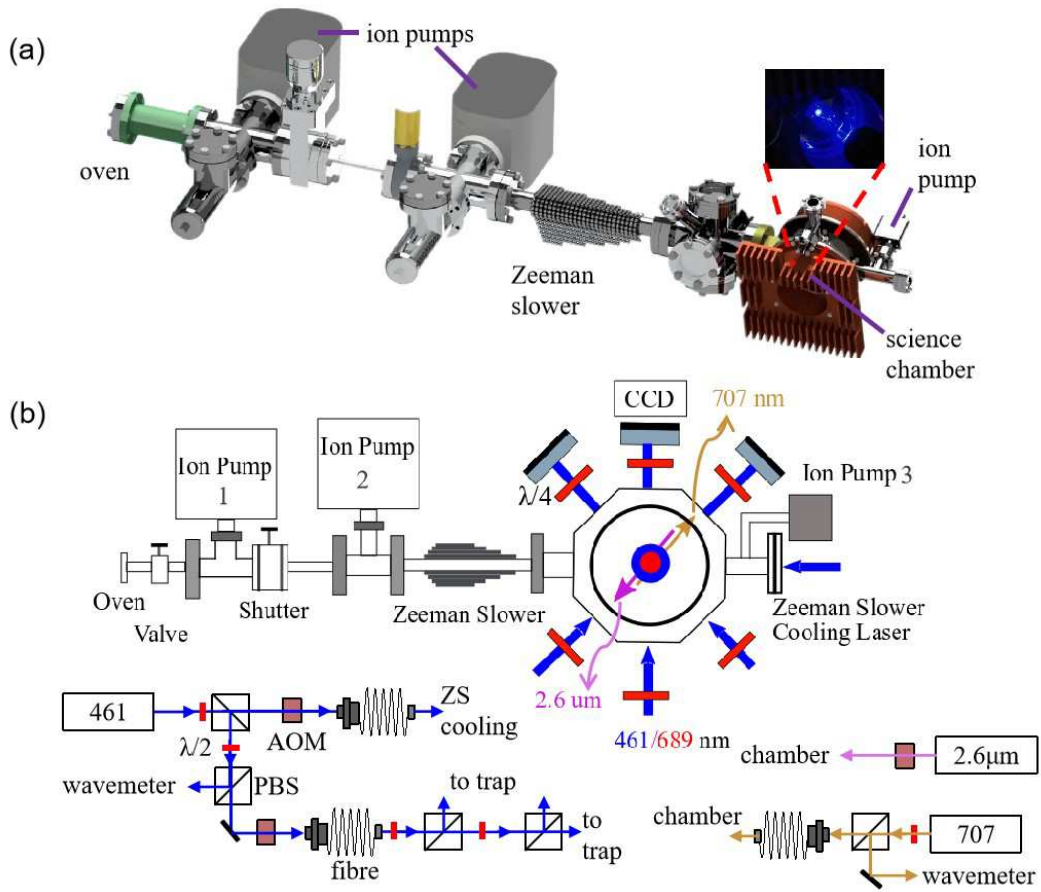}% Here is how to import EPS art
\caption{\label{fig:epsart} The experimental setup for cooling and trapping Sr atoms. (a) CAD drawing of the setup consisting of self-assembled Zeeman slower and the science chamber. Inset: a typical image of a blue MOT consisting of 9.8$\times10^{8}$ atoms. (b) Schematic diagram of the experimental setup including the plan of cooling and trapping laser beams. Blue arrows represent 461$\,$nm laser beams; The directions of 707$\,$nm and 2.6$\,\mu$m beams are pointed inward and outward, respectively; CCD, camera for taking images of atoms; $\lambda$/2, half-wave plate; AOM, acousto-optic modulator; PBS,  polarization beam splitter; ZS, Zeeman slower.}
\end{figure*}
%The spin-forbidden $^{1}$S$_{0}$-$^{3}$P$_{1}$ 689 nm transition with a 7.6 kHz linewidth and the 3 $\mu$W/cm$^{2}$ saturated intensity is used in the second stage cooling. This transition is so narrow that the red laser can only cover less than 2$\%$ of the Doppler broadening corresponding to the velocity distribution of the atoms in blue MOT. Therefore, the laser needs initially to be broadened to capture the maximum number of atoms and then gradually decrease the broadening until the minimum temperature is reached. The initial phase is called broadband red MOT and the second phase is single frequency red MOT. A modulation here will be added to the red laser to broaden the spectrum of of the red laser. The temperature of cold atoms can be cooled low to a few $\mu$K in the red MOT. The recoil temperature is 490 nK limited by the recoil energy of photons at this wavelength[recoil].
We employ the standard six-beam MOT of alkaline earth atoms as the experimental setup for cooling Sr atoms, shown in Fig. 2. Two ion pumps of 25$\,$L$s^{-1}$ each placed close to the atomic source prevent the pressure around the oven from rising when in operation. A self-assembled Zeeman slower placed after the oven provides a high-flux atomic beam with a significant portion of atoms within the capture velocity of the MOT of 70$\,$m$s^{-1}$. The science chamber is a custom made spherical octagon made of titanium which has a side length of 35 mm and depth of 36 mm. The third ion pump with a pumping speed of 2.5$\,$L$s^{-1}$ is added near the science chamber to further improve the vacuum. We achieve a pressure of 1$\times$10$^{-11}$ mbar inside our science chamber. A CCD camera (Andor Zyla 5.5) is used to detect fluorescence from MOT.

%A self-assembled Zeeman slower, described in section \ref{sec:level2}, placed after the oven provides a high-flux atomic beam with a greater portion of atoms within the capture velocity of the MOT than the thermal atoms released by the oven. The science chamber is a custom made spherical octagon made of titanium which has a side length of 35$\,$mm and depth of 36$\,$mm. The third ion pump with a pumping speed of 2.5$\,$Ls$^{-1}$ is added near the science chamber to further improve the vacuum. We achieve a pressure of 1$\times10^{-11}$$\,$mbar inside our science chamber. A CCD camera (Andor Zyla 5.5) is used to detect fluorescence from MOT.

The cooling and trapping light at 461$\,$nm, generated by frequency doubling 922$\,$nm light from a SolstiS Ti:Sapphire laser and doubling unit from M squared lasers, has a total power of 550$\,$mW at 461$\,$nm. A small sample of the output is coupled to an optical fibre for frequency stabilisation and monitoring with a wavemeter (WSU2 from highfinesse). The wavemeter has a resolution of 2$\,$MHz and is calibrated by a SLR-780 rubidium reference laser. The main output of the 461$\,$nm laser is divided into three branches: the Zeeman slower beam, the probe beam and the MOT beam. All the three beams are coupled to an optical system including acousto-optical modulators (AOMs), polarization beam splitters (PBSs), waveplates, fiber couplers, etc, delivering light to the experiment. The laser power can be precisely controlled by AOMs driven by a direct digital synthesizer (model EVAL-AD9959). The MOT beams are expanded to 1$\,$cm in radius and the slower beam is focused to be 1.1$\,$mm at the oven nozzle.

We employ $^{3}$P$_{2}$$\rightarrow$$^{3}$S$_{1}$ at 707$\,$nm and $^{3}$P$_{0}$$\rightarrow$$^{3}$D$_{1}$ at 2.6$\,\mu$m instead of $^{3}$P$_{0}$$\rightarrow$$^{3}$S$_{1}$ at 679$\,$nm as repumpers. The two repump lasers of 707$\,$nm and 2.6$\,\mu$m are home-built with a power of 2.5$\,$mW and 7$\,$mW, respectively. The 707$\,$nm laser is locked to the wavemeter while 2.6$\,\mu$m laser is free running at this stage.
%The 689 nm cooling and trapping laser is a master diode laser and an injection locked slave laser. To provide the stable laser output, the laser is -100 MHz shifted and locked to an ultra low expansion cavity placed in a temperature stabilized vacuum chamber with a pressure of 9$\times$10$^{-8}$mBar. The cavity has a finesse of 300,000 and a free spectral range of 1.5 GHz. Another AOM is used to bridge the frequency offset between the laser frequency and the atomic resonance. A frequency modulated radio frequency signal is used to drive this AOM. 

A Zeeman slower is required for the preparation of a high-flux source of cold Sr. Our Zeeman slower based on Ref\cite{Selfassembled} is a permanent magnet Zeeman slower based on spherical NdFeB magnets. It is a longitudinal-field Zeeman slower and uses only $\sigma^{+}$ or $\sigma^{-}$ polarized light. The slower is mechanically stable, quick to assemble, and doesn't need a mechanical holder. Atoms with the velocity between 200$\,$ms$^{-1}$ and 350$\,$ms$^{-1}$ can be slowed down by slower and captured in the MOT. The most probable velocity of atomic beam is reduced from 400$\,$ms$^{-1}$ to 94$\,$ms$^{-1}$ by slower, and the velocity width is narrowed from 370$\,$ms$^{-1}$ to 26$\,$ms$^{-1}$. The capture efficiency of slower is 34$\%$.

\section{Novel repumping scheme}
By heating the oven up to 560$^{\circ}$C, it effuses an atomic beam with a high flux of 2.8$\times$10$^{11}\,$s$^{-1}$. Under the typical condition, the combined MOT beams at 461$\,$nm have an intensity of 19$\,$mW/cm$^{2}$ ($\sim$0.5I$_{s}$), 1/e$^{2}$ radius of 10$\,$mm and a detuning of -40$\,$MHz (1.25$\Gamma$). The magnetic field gradient is 55$\,$G/cm. The experimental loading processes of blue MOT in different cases are shown in Fig. 3. The atom number of 2$\times$10$^{6}$ is collected in the MOT without Zeeman slowing and repumping beams. When the slowing beam and 707$\,$nm and 2.6$\,\mu$m repumpers are turned on, the steady-state atom number is enhanced by a factor of 15 with respect to the non-repumping case, while the factor is around 11 in the case of 679 nm repumper replacing the 2.6$\,\mu$m repumper, resulting in an enhancement ratio of 1.4 between them. Due to the uncertainty of atom number measurements, the above ratio could range from 1.3 to 1.5. The laser intensity of both 2.6$\,\mu$m and 679$\,$nm lasers is 20$\,$mW/cm$^{2}$ while in operation. 

\begin{figure}[ht]
%\begin{minipage}{6.3cm}
\includegraphics[scale=0.38]{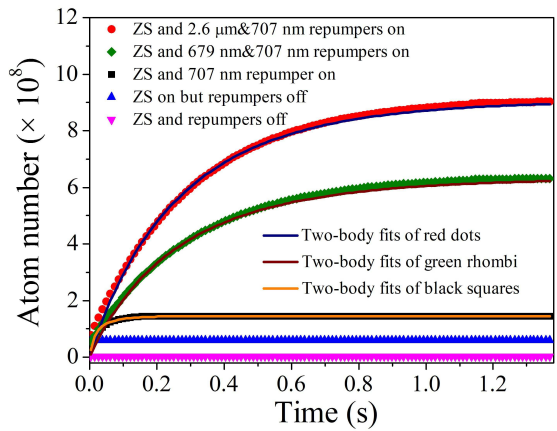}% Here is how to import EPS art
\caption{\label{fig:epsart} The dependence of atom number in blue MOT on Zeeman slower beam and two repumping beams. The solid curves represent our model fits. All the dots with different colors represent the experimental data.}
\end{figure}

To exam the enhancement of steady-state MOT atom number induced by repumping lasers, we use time-dependent trap loading equation \begin{equation}
\dot{N}=L-\Gamma N-\beta'N^{2}
\end{equation}
Here, $N$ is atom number in the MOT, $L(=3\times10^{9}$s$^{-1})$ is MOT loading rate, $\Gamma$ is one-body loss rate, $\beta'=\beta/(2\sqrt{2}V)$, where $\beta$ is two-body loss constant and $V$ is an effective volume for two-body processes. The solution to this differential equation is
\begin{equation}
N(t)=\frac{N^{ss}(1-e^{-\gamma t})}{1+\chi e^{-\gamma t}}
\end{equation}
where $N^{ss}$ is the steady-state MOT atom number, $\gamma$ represents the total loss rate and $\chi$ is relative contributions of
the one- and two-body loss coefficients. They are given by 
\begin{equation}
\gamma=\Gamma+2\beta'N^{ss}
\end{equation}
\begin{equation}
\chi=\frac{\beta'N^{ss}}{\beta'N^{ss}+\Gamma}
\end{equation}

We use this model to fit the experimental data of three different repumping schemes shown in Fig. 3. One- and two-body loss rates can be extracted from the fits. The case of 2.6$\,\mu$m is fitted by the two-body fit with $\Gamma=2.54\pm0.05\,$s$^{-1}$ and $\beta=6.40\pm0.34\times10^{-12}\,$cm$^{3}$/s. In the case of 679$\,$nm scheme, $\Gamma=3.03\pm0.05\,$s$^{-1}$ and $\beta=8.89\pm0.34\times10^{-12}\,$cm$^{3}$/s. The results indicate that the relatively smaller one-body loss rate in 2.6$\,\mu$m case leads to an enhancement of the steady-state atom number, in comparison with 679$\,$nm case.

Next, to find out the dependence of steady-state atom number on the intensities of repumpers, a set of rate equations is used. This has been applied in single-laser repumping schemes\cite{Stellmer,FUYU,Teixei}. Here, we extend it to a dual repumping scheme, i.e. 707$\,$nm and 2.6$\,\mu$m repumpers in our case. There are eight states involved in our case which are labeled with numbers shown in Fig. 1. The rate equations of these states are given by
\begin{equation}
\begin{aligned}
&\dot{N_{1}}=L+B(N_{2}-N_{1})+\Gamma_{21}N_{2}+\Gamma_{71}N_{7}\\
&\dot{N_{2}}=-B(N_{2}-N_{1})-(\Gamma_{21}+\Gamma_{23})N_{2}\\
&\dot{N_{3}}=\Gamma_{23}N_{2}-(\Gamma_{34}+\Gamma_{37})N_{3}\\
&\dot{N_{4}}=\theta(N_{5}-N_{4})+\Gamma_{34}N_{3}+\Gamma_{54}N_{5}+\Gamma_{64}N_{6}-\Gamma_{d}N_{4}\\
&\dot{N_{5}}=-\theta(N_{5}-N_{4})-(\Gamma_{54}+\Gamma_{57}+\Gamma_{58}-\Gamma_{d})N_{5}\\
&\dot{N_{6}}=-\alpha(N_{6}-N_{8})-(\Gamma_{64}+\Gamma_{67}+\Gamma_{68}-\Gamma_{d})N_{6}\\
&\dot{N_{7}}=\Gamma_{37}N_{3}+\Gamma_{57}N_{5}+\Gamma_{67}N_{6}-\Gamma_{71}N_{7}\\
&\dot{N_{8}}=\alpha(N_{6}-N_{8})+\Gamma_{58}N_{5}+\Gamma_{68}N_{6}-\Gamma_{d}N_{8}
\end{aligned}
\end{equation}
Where $\Gamma_{ij}$ is transition rates from $i$ to $j$. $B$, $\theta$ and $\alpha$ are pumping rates of the 461$\,$nm, 707$\,$nm and 2.6$\,\mu$m transition, respectively, which are proportional to their respective laser intensities. The intensity of 19$\,$mW/cm$^{2}$ for 461$\,$nm laser results in $B=3.8\times10^{7}$s$^{-1}$. The loss rate of $\Gamma_{d}$ is added to all levels involved in repumping that are dark for the cooling light\cite{Teixei}. The value for which can be deduced from the fit to experimental data. All the transition rates and branching ratios of these states are summarized in Table~\ref{tab:table1}. By solving the equations, we obtain the dependence of steady-state atom number on the pumping rates of $\alpha$ and $\theta$. This is plotted in Fig. 4(a). The atom number increases as the pumping rates of 707$\,$nm and 2.6$\,\mu$m lasers increase. The blue area shows that the atom number is very low when the laser intensity of either 707$\,$nm or 2.6$\,\mu$m is small. The atom number is the same along the contour lines.
 
We apply the same rate equations to the 679$\,$nm case to make a comparison with 2.6$\,\mu$m case. The only difference in 679$\,$nm case is that there is no |6> involved. Thus the rate equations of |4>, |5>, |7> and |8> are revised as follows,
\begin{equation}
\begin{aligned}
\dot{N_{4}}=&\theta(N_{5}-N_{4})+\Gamma_{34}N_{3}+\Gamma_{54}N_{5}-\Gamma_{d}N_{4}\\
\dot{N_{5}}=&-\theta(N_{5}-N_{4})-\alpha(N_{5}-N_{8})-(\Gamma_{54}+\Gamma_{57}\\
&+\Gamma_{58}-\Gamma_{d})N_{5}\\
\dot{N_{7}}=&\Gamma_{37}N_{3}+\Gamma_{57}N_{5}-\Gamma_{71}N_{7}\\
\dot{N_{8}}=&\alpha(N_{5}-N_{8})+\Gamma_{58}N_{5}-\Gamma_{d}N_{8}
\end{aligned}
\end{equation}

The steady-state MOT atom number for 2.6$\,\mu$m (blue) and 679$\,$nm (red) cases as a function of pumping rate of $\alpha$ at different pumping rates of $\theta$ are plotted in Fig. 4(b)-(d). Other parameters are kept the same for both cases. The atom number in both cases increases and reaches a saturation value as a function of $\alpha$ with $\theta$ being fixed. When 2.6$\,\mu$m and 679$\,$nm lasers are working at 20$\,$mW/cm$^{2}$, their corresponding pumping rates are 5.3$\times10^{9}$s$^{-1}$ and 1.8$\times10^{7}$s$^{-1}$. Utilizing the experimental parameters, we have simulated atom numbers for both cases. These are shown by stars in Fig. 4(c). The ratio between the two simulated atom numbers is calculated to be 2, which is in a reasonably good agreement with the averaged experimental ratio of 1.4. The reason for the discrepancy is due to the uncertainty of 4$\%$ in the atom number measurement. 

%In this paper, we are sensitive to the relative repumping efficiency of both repumping schemes. To compare the repumping effect between these two schemes, all parameters in 679$\,$nm case are kept as the same as 2.6$\,\mu$m case. Under the same laser intensity, the pumping rate $\alpha$ of 679$\,$nm is 1.8$\times$10$^{7}\,$s$^{-1}$. The relative enhancement can be present by the ratio($R$) of enhancements in the two schemes, given by
%\begin{equation}
%R=\frac{(\Gamma_{57}\Gamma_{64}+\Gamma_{57}\Gamma_{67}+\Gamma_{58}\Gamma_{67})(\Gamma_{54}+\Gamma_{57}+\beta)}{\Gamma_{57}(\Gamma_{64}+\Gamma_{67})(\Gamma_{5}+\beta)}
%\end{equation}
%It only depends on the transition rates $\Gamma_{5,6j}$ and $\beta$. The simulated results of the model are shown in Fig. 4. Here the relation between $\beta$ and $s$ on resonance is 
%\begin{equation}
%\beta=\frac{6\pi c^{2}\Gamma_{54}I_{s}}{\hbar\omega^{3}\Gamma_{5}}s
%\end{equation}
%$I_{s}$ and $s$ are the saturation intensity and saturation parameter of 707$\,$nm laser. When the value of $\alpha$ is given, the enhancement of atom number is increased as the intensity of 707$\,$nm laser increases for both schemes, and reach a saturation at $s=\sim1000$. The enhancement in 2.6$\,\mu$m case is larger than that of 679$\,$nm case, and the relative enhancement ratio is between 1.22 and 1.38.
\begin{table}
\caption{\label{tab:table1}%
Transition rates ($\Gamma$), branching ratios for relevant transitions of Sr. Data are referred from Refs\cite{Stellmer,Porseand}.}
\begin{ruledtabular}
\begin{tabular}{ccc}
%\textrm{Transition\footnote{Note a.}}&
%\textrm{Transition}&
\multicolumn{1}{c}{\textrm{Transition}}&
\multicolumn{1}{c}{\textrm{Transition rate ($\times$10$^{5}$$s^{-1}$)}}&
\textrm{Branching ratio ($\%$)}\\
\colrule
$^{1}$P$_{1}$$\rightarrow$$^{1}$S$_{0}$ & 2000 & -\\
$^{1}$P$_{1}$$\rightarrow$$^{1}$D$_{2}$ & 0.039 & 0.002\\
$^{1}$D$_{2}$$\rightarrow$$^{3}$P$_{2}$ & 0.0066 & 33.3\\
$^{1}$D$_{2}$$\rightarrow$$^{3}$P$_{1}$ & 0.0134 & 66.7\\
$^{3}$S$_{1}$$\rightarrow$$^{3}$P$_{2}$ & 400 & 55.6\\
$^{3}$S$_{1}$$\rightarrow$$^{3}$P$_{1}$ & 260 & 33.3\\
$^{3}$S$_{1}$$\rightarrow$$^{3}$P$_{0}$ & 87 & 11.1\\
$^{3}$D$_{1}$$\rightarrow$$^{3}$P$_{2}$ & 0.088 & 1.9\\
$^{3}$D$_{1}$$\rightarrow$$^{3}$P$_{1}$ & 1.8 & 38.7\\
$^{3}$D$_{1}$$\rightarrow$$^{3}$P$_{0}$ & 2.8 & 59.4\\
$^{3}$P$_{1}$$\rightarrow$$^{1}$S$_{0}$ & 0.47 & -\\
\end{tabular}
\end{ruledtabular}
\end{table}

%The atom number of a continuously repumped MOT is dependent on both the repumping light intensities in dual repumping schemes. To compare the relative repumping efficiency of 2.6$\,\mu$m and 679$\,$nm repumpers, we fix the intensity at 16$\,$mW/cm$^{2}$ for 2.6$\,\mu$m and 679$\,$nm lasers, corresponding to 3$\times$10$^{4}$ and 27 times their respective saturation intensities. To simplify the calculation of equations, we assume $\Gamma_{4x}$$\gg$$\Gamma_{5,6,8x}$. 

%In our case, the 707$\,$nm repumping intensity is approximately 10 times higher than its saturation intensity (I$_{s}$=2.3$\,$mW/cm$^{2}$), and hence $\beta\approx$120$\,$MHz. 

%Moreover, the enhanced enhancement of atom number for 2.6$\,\mu$m case can also be roughly elucidated by the repumping efficiency calculated from the branching ratios. The repumping efficiency for 2.6$\,\mu$m case is 95.3$\%$, much larger than 37.5$\%$ for 679$\,$nm case.

\begin{figure*}
\includegraphics[scale=0.7]{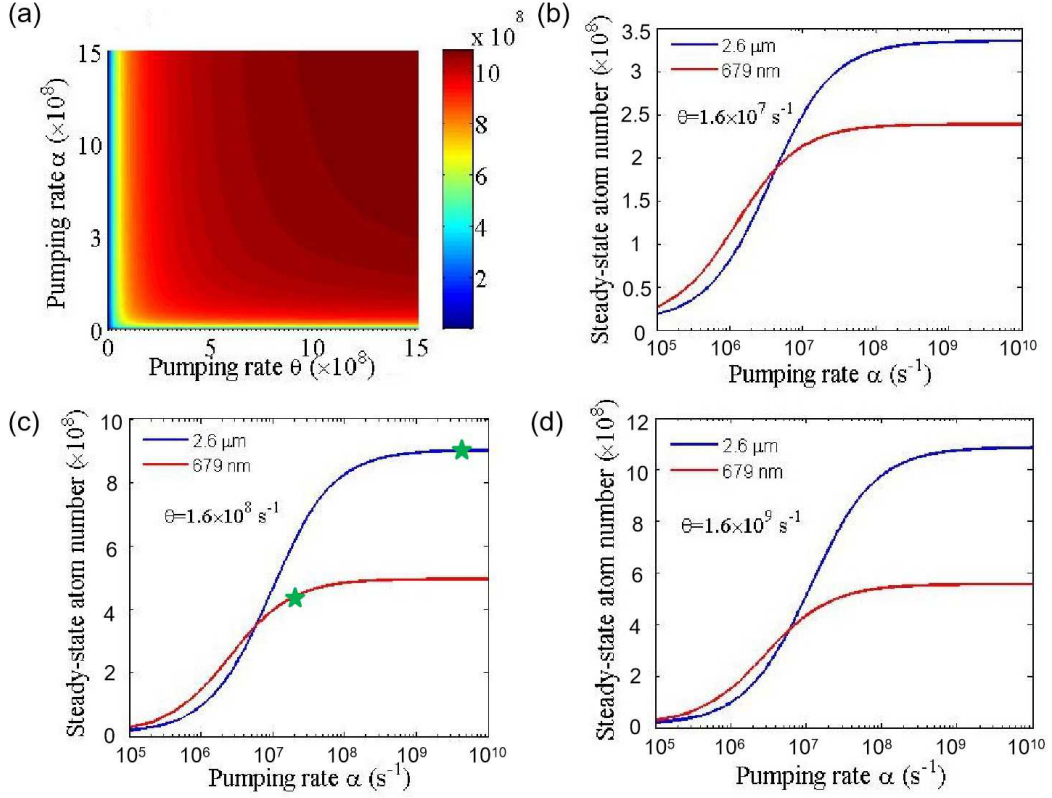}% Here is how to import EPS art
\caption{(a) The dependence of steady-state MOT atom number on pumping rates of $\alpha$ and $\theta$. (b), (c), (d) show the steady-state MOT atom number as a function of pumping rate of $\alpha$ for 2.6$\,\mu$m (blue curve) and 679$\,$nm (red curve) cases when $\theta=1.6\times10^{7}$s$^{-1}$, $\theta=1.6\times10^{8}$s$^{-1}$, $\theta=1.6\times10^{9}$s$^{-1}$, respectively.}
\end{figure*}
%To simplify the calculation, we assume $\Gamma_{5x}$$\ll$$\Gamma_{5j}$ and $\Gamma_{6x}$$\ll$$\Gamma_{6j}$. The enhancement could be written as 
%\begin{equation}
%\varepsilon=\frac{\beta(\Gamma_{57}\Gamma_{64}+\Gamma_{57}\Gamma_{%67}+\Gamma_{58}\Gamma_{67})}{\Gamma_{4x}(\Gamma_{64}+\Gamma_{67})(\Gamma_{5}+\beta)}
%\end{equation}
%Where $\Gamma_{5}$ denotes the total decay rate, given by $\Gamma_{5}=\Gamma_{54}+\Gamma_{57}+\Gamma_{58}$=74.7 MHz.

%Table~\ref{tab:table1},%
%\begin{table}[b]%The best place to locate the table environment is directly after %its first reference in text

%The time-dependent number equation for MOT loading:
%\begin{equation}
%\dot{N}=L-\Gamma N-\beta'N^{2}
%\end{equation}
%Here, $N$ is the atom number in the MOT, $\Gamma$ is the one-body loss rate, $\beta'=\beta/(2\sqrt{2}V)$, where $\beta$ is the two-body loss constant. The solution to this differential equation is
%\begin{equation}
%N(t)=\frac{N^{ss}(1-e^{-\gamma t})}{1+\chi e^{-\gamma t})}
%\end{equation}
%with $\gamma=\Gamma+2\beta'N^{ss}$ and 
%\begin{equation}
%\chi=\frac{\beta'N^{ss}}{\beta'N^{ss}+\Gamma}
%\end{equation}
%Using this model, we determine the fits shown in Fig. 4. When only having 707$\,$nm repumper, a two-body fit fits the data with $\Gamma=3.9\,s^{-1}$, $\beta=3.8\times10^{-10}\,$cm$^{3}/s$. When the 2.6$\,\mu$m repumper is added, the two-body fit indicates $\Gamma=2.0\,s^{-1}$, $\beta=1.0\times10^{-11}\,$cm$^{3}/s$. Comparatively, the 679$\,$nm case has $\Gamma=2.3\,s^{-1}$, $\beta=2.1\times10^{-11}\,$cm$^{3}/s$ by two-body fit.

\section{Characterization of our MOT}

We apply the dual repumping scheme of 707$\,$nm and 2.6$\,\mu$m transitions, together with the self-assembled Zeeman slower, to increase the atom number in a continuously repumped MOT. By optimizing the magnetic field, laser powers, polarizabilities, detunings of MOT and slowing beams, at the oven temperature of 560$^{\circ}$C, 9.8$\times$10$^{8}$ atoms (1.4$\times$10$^{11}\,$cm$^{-3}$ for atom density) are collected in the continuously repumped MOT. This number is comparable to 8$\times$10$^{8}$ at 600$^{\circ}$C and 1.3$\times$10$^{9}$ at 630$^{\circ}$C, previously achieved in Ref\cite{Courti03}. However, the oven temperature in our system is lower. We can enhance the atom number up to 1.5$\times$10$^{9}$ by heating the oven to 600$^{\circ}$C. Furthermore, the blue MOT temperature is measured using the time of fight (TOF) method\cite{TOF}. The atoms are released from the trap and allowed to freely expand while they fall under gravity. The TOF images are taken by the CCD camera. The images at different TOF are shown in inset of Fig. 5. The measured temperature is 1.1(2)$\,$mK.
\begin{figure}[ht]
%\begin{minipage}{6.3cm}
\includegraphics[scale=0.38]{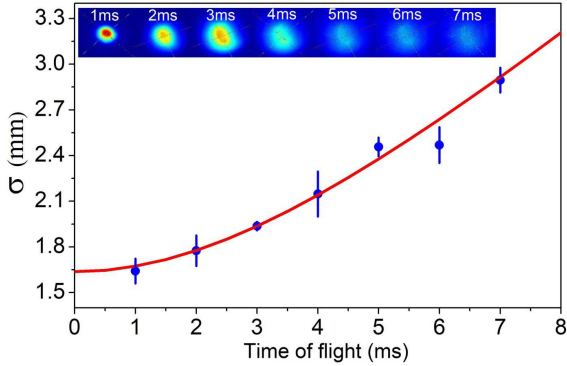}% Here is how to import EPS art
\caption{\label{fig:epsart} The radius of blue MOT as a function of time of flight. Inset: the images of atomic cloud at different time of flight.}
\end{figure}

In order to further characterize the setup, the lifetime of $^{3}$P$_{2}$ magnetic trap is measured. The details are as follows. Firstly, atoms are continuously populated into $^{3}$P$_{2}$ during the blue MOT process for 1.3$\,$s loading time. Secondly, 461$\,$nm laser is switched off but the magnetic field is held on for $t_{hold}$ which can be varied. Finally, switch off magnetic field and switch on 707$\,$nm laser, 461$\,$nm probe laser and CCD camera after $t_{hold}$. $^{3}$P$_{2}$ magnetic trap atoms are detected by collecting the fluorescence. 2.6$\,\mu$m repumping laser is kept on throughout. The corresponding sequence is shown in Fig. 6(a). The magnetic field gradient of 50$\,$G/cm gives a magnetic trap depth of 38$\,$mK. The measurements are shown with blue dots in Fig. 6(a). It indicates the lifetime of the magnetic trap to be 1.1$\,$s, which is comparable to the Ref\cite{Nagel}.
\begin{figure}[ht]
%\begin{minipage}{6.3cm}
\includegraphics[scale=0.37]{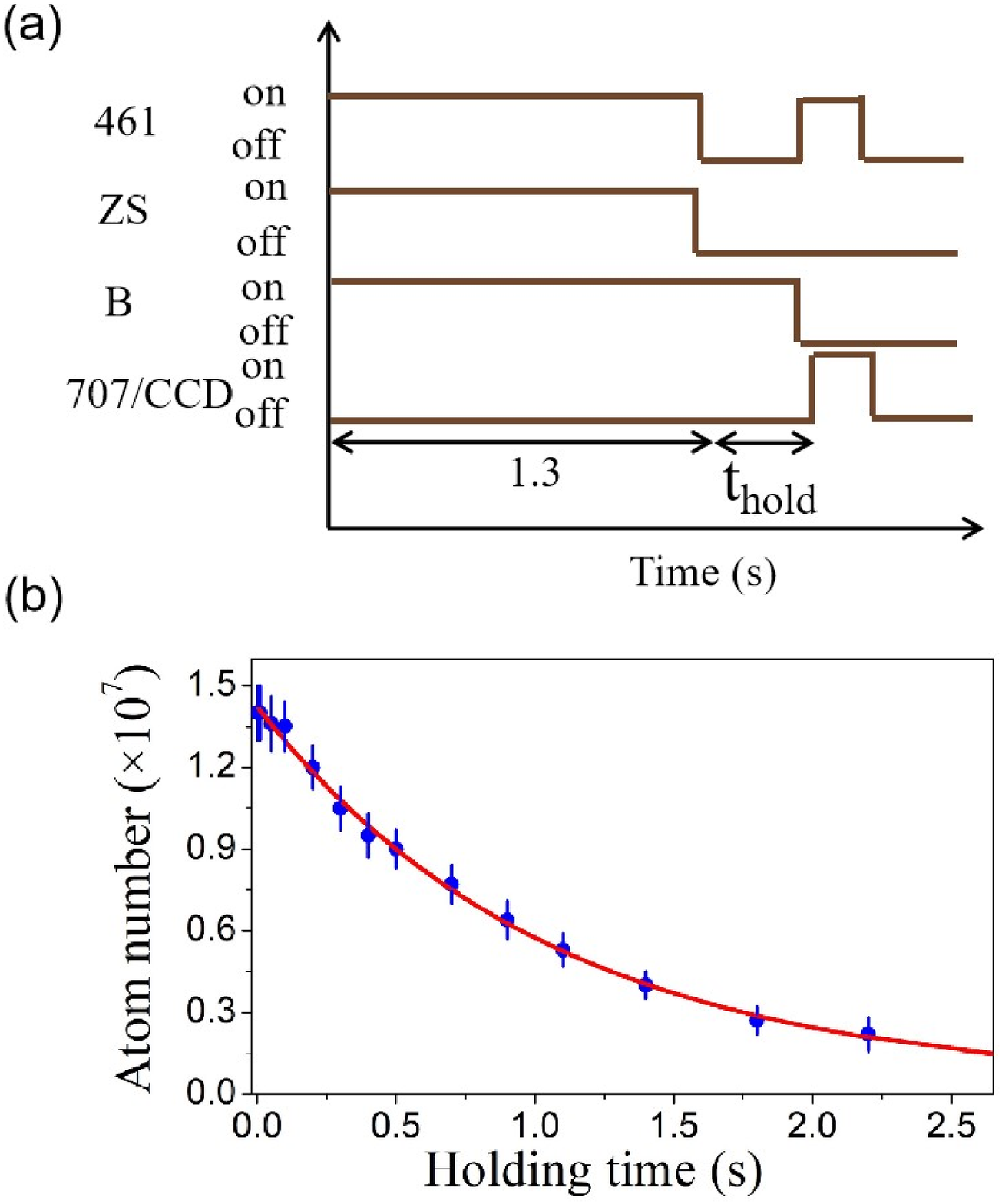}% Here is how to import EPS art
\caption{\label{fig:epsart} (a) The magnetic-trap decay timing sequence. ZS: Zeeman slower; B: magnetic field gradient; CCD: charge coupled device camera. In the lifetime measurement, the atoms are continuously loaded into $^{3}$P$_{2}$ for 1300 ms. (b) Lifetime measurement of $^{3}$P$_{2}$ magnetic trap (blue solid circles). Fit is shown by the red curve. It indicates that the lifetime of the magnetic trap is 1.1 s.}
\end{figure}

\section{$^{3}$D$_{1}$ Land\'{E} g factor Measurement}
%There are two main reasons to measure the $Land\acute{e}$ g factor of $^{3}$D$_{1}$ of $^{88}$Sr. 
In alkaline-earth-like atoms, the $g$ factors of $^{1}$S$_{0}$ and $^{3}$P$_{0,1,2}$ have been experimentally determined to estimate the Zeeman shift in optical clocks\cite{NuBoyd}. Here, we measure the Land\'{e} $g$ factor of $^{3}$D$_{1}$ using cold $^{88}$Sr atoms trapped in a blue MOT.

The resolved-sideband Zeeman spectroscopy is the key step for the $g$ factor measurement. $^{88}$Sr atoms are continuously loaded into the blue MOT with the 2.6$\,\mu$m and 707$\,$nm repump lasers on throughout. A neutral density filter is inserted into the reflected MOT beam parallel to the anti-Helmhotz coil axis, and hence resulting an imbalance between the traveling-wave components. The cloud is pushed towards the reflected beam side due to the intensity imbalance. The 2.6$\,\mu$m laser is frequency scanned to cover the splittings. Atoms excited to $^{3}$D$_{1}$ by the resonant 2.6$\,\mu$m laser will decay back to the 461$\,$nm transition cooling cycle via $^{3}$P$_{1}$ to the ground state. Through detecting the blue fluorescence with a photomultiplier tube (PMT), Zeeman spectroscopy of $^{3}$P$_{0}\rightarrow^{3}$D$_{1}$ is obtained while scanning 2.6$\,\mu$m laser frequency. The procedure is repeated with different filters inserted in the same position. The experimental parameters of 2.6$\,\mu$m laser are 80$\,\mu$W power and 2.3$\,$mm beam size. The scanning speed is set at 8$\,$MHz/s and the magnetic field gradient in operation is 45$\,$G/cm.  

Fig. 7 shows the typical resolved-sideband Zeeman spectroscopy of $^{3}$P$_{0}$$\rightarrow$$^{3}$D$_{1}$. The three peaks correspond to m = 0$\rightarrow$m' = -1, 0, +1 from left to right. Here, a NE02A filter with OD = 0.2 is inserted which can produce 37$\%$ intensity difference. All the three peaks are fitted with a Lorentz function. They are broadened by three independent effects: (i) the intensity broadening, (ii) the linewidth of 2.6$\,\mu$m laser and (iii) the Doppler broadening. The side peaks are additionally broadened by the magnetic field variation across the atomic cloud. The Doppler broadening is estimated to be 175$\,$kHz at the temperature of 1.1$\,$mK. The intensity broadening can be calculated as 4.4$\,$MHz by the relationship $\Delta\nu = \Delta\nu_{N}\sqrt{1+I/I_{S}}$, where $\Delta\nu_{N}$ is the natural linewidth of $^{3}$D$_{1}$, $I_{S}$ is the saturated intensity for $^{3}$P$_{0}$$\rightarrow$$^{3}$D$_{1}$. The full width at half maximum (FWHM) of the three peaks are 18.3$\,$MHz, 6.3$\,$MHz and 14.7$\,$MHz from left to right. The frequency shifts of the left and right side peaks relative to the central peak are 10.2$\,$MHz and 10.8$\,$MHz, respectively. The measured difference between the left and right peaks in FWHM and shift is mainly caused by the fitting uncertainty and the induced magnetic field variation by the atomic cloud within a laser scanning period. 
\begin{figure}[ht]
\includegraphics[scale=0.35]{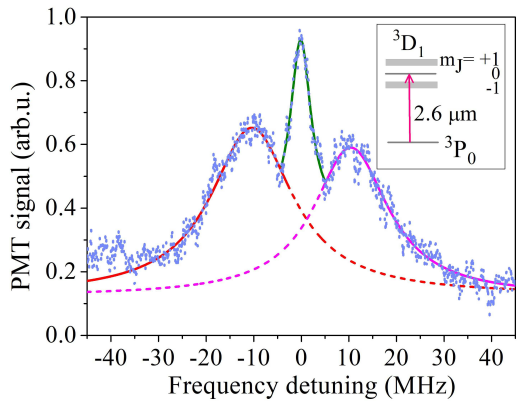}% Here is how to import EPS art
\caption{\label{fig:epsart}A typical resolved-sideband Zeeman spectroscopy of $^{3}$P$_{0}$$\rightarrow$$^{3}$D$_{1}$. Experimental data (blue dots) are fitted with three independent Lorentz functions (colourful curves). The three peaks from left to right correspond to $m$ = 0$\rightarrow$$m'$ = -1, 0, +1 transitions. Inset: the Zeeman sublevels of $^{3}$D$_{1}$.}
\end{figure}
%\begin{equation}
%\end{equation}

To validate the experimental calibration of peak splittings and widths, we calculate them by the displacement ($S$) and atomic cloud radius ($\Delta S/2$)\cite{probe},
\begin{equation}
S=\frac{\Delta_{S}h}{g\mu_{B}m_{J}B^{'}_{Z}},
\end{equation}
and
\begin{equation}
\frac{\Delta S}{2}=\frac{\Delta_{W}h}{g\mu_{B}m_{J}B^{'}_{Z}},
\end{equation}
where $\Delta_S$ and $\Delta_{W}$ are the peak splitting and the FWHM difference between the central peak and side peaks. $\mu_{B}$ is the Bohr magneton, $m_{J}=\pm$1 is the magnetic quantum number, and $h$ is Planck's constant. For the case of the OD=0.2 
filter and the data shown in Fig. 7, $S = 2.98\,$mm and $\Delta S/2= 2.6\,$mm, which are transferred from the unit in pixel by 38$\,\mu$m/pixel. According to equations (7) and (8), $\Delta_S$ and $\Delta_{W}$ are calculated as 9.4$\,$MHz and 8.2$\,$MHz, respectively, which are consistent with the experimental values of 10.5$\,$MHz and 10.2$\,$MHz. The slight increase of measured results is believed to arise from the experienced higher magnetic field by the atomic cloud in the experiment.
\begin{figure}[ht]
\includegraphics[scale=0.3]{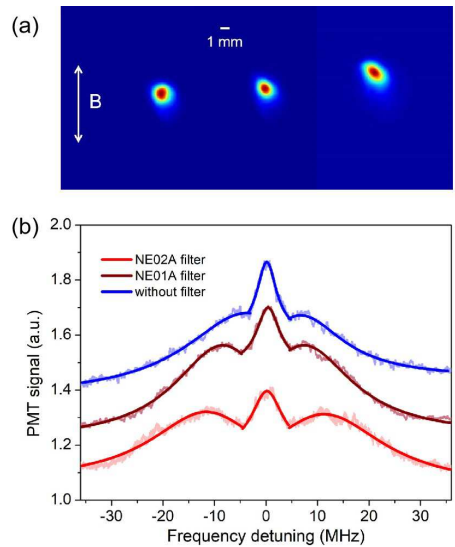}
\caption{\label{fig:epsart}(a) The atomic clouds from left to right are located in the center of MOT (without filter), at a position of small magnetic field (with a NE01A filter) and higher magnetic field (with a NE02A filter), respectively. The cloud moves towards the reflected beam side along the magnetic field direction. (b) The Zeeman spectroscopy in the above three cases. Light dots represent experimental data and dark curves are fitted results for all three cases. The splitting of the Zeeman sublevels becomes larger and the amplitude of the signal is weaker as the magnetic field increases.}
\end{figure}
\begin{figure}[htbp]
\includegraphics[scale=0.36]{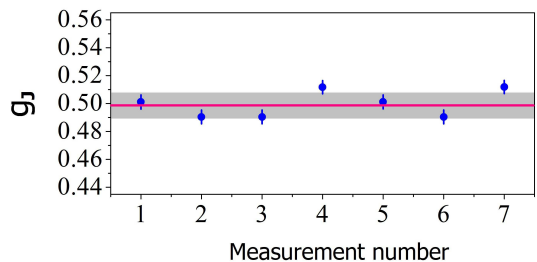}% Here is how to import EPS art
\caption{\label{fig:epsart}Summary of measurements of the Land\'{e} $g$ factor of $^{3}$D$_{1}$. The error bars show the uncertainty of the experiment. The soild line and the gray area represent the mean measured value and the 1-$\sigma$ confidence interval.}
\end{figure}

The Zeeman spectroscopy of $^{3}$P$_{0}$$\rightarrow$$^{3}$D$_{1}$ at different magnetic fields are detected to experimentally measure $g$ value. In Fig. 8, we plot the Zeeman spectroscopy in three different field cases, i.e., no filter, NE01A filter and NE02A filter with intensity differences of 0$\%$, 21$\%$ and 37$\%$. As the intensity difference increases, the atomic cloud moves to the higher magnetic field region as shown in Fig. 8(a), and the $m_{J} = \pm$1 levels are split more. Thus the two side peaks of Zeeman spectroscopy are more separated and broadened shown in Fig. 8(b). It's notable that all three peaks in the case of no filter are not completely overlapped, which is due to the spatial distribution of the cloud. Each case is repeated by 7 times. By the relation of $g = \Delta_{S} h/(\mu_{B}B)$, we can deduce the value of $g$ factor as the seven dots shown in Fig. 9. The pink solid line means the mean value 0.4995 of these seven numbers, and the gray area represents 1-$\sigma$ confidence interval. The statistical error in $g$ measurement is significantly contributed by the uncertainty in the magnetic field measurement caused by inhomogeneity of quadruple magnetic field, imperfect optical path alignments and residual fields. 
%The Hamiltonian in the external magnetic field $B$ is given by\cite{gJ}
%\begin{equation}
%H=H_{0} + AL\cdot S + g_{J}\mu_{B}J\cdot B + e^2/8m\sum_{a}(B \times  r_{a})^2
%\end{equation}
%where the second term takes into account the fine structure coupling, the third term is the Zeeman effect, and the last term is the diamagnetic correction. 

To compare the measured $g$ value with theoretically calculated one, we calculate it by the common Russel-Saunders approximation\cite{gJ},
\begin{equation}
\begin{aligned}
g_{J}=g_{L}\frac{J(J+1)+L(L+1)-S(S+1)}{2J(J+1)}, \\
+ g_{S}\frac{J(J+1)+S(S+1)-L(L+1)}{2J(J+1)},
\end{aligned}
\end{equation}
where $L$ is the total orbital momentum quantum number, $S$ is the total spin quantum number, $J$ is the total electronic angular momentum quantum number. Here, $g_{L}$ = 1, $g_{S}$ = 2 $\times$ 1.0011597\cite{1986}. The value of $g_{J}$ is calculated to be 0.4988 for $^{3}$D$_{1}$ state of $^{88}$Sr, which is in good agreement with our mean measured value 0.4995. 

\section{conclusion}
In conclusion, we have presented an experimental setup for cooling and trapping of Sr atoms. The atom number trapped in the blue MOT can be enhanced by up to a factor of 15 with the use of a self-assembled Zeeman slower based on permanent magnets and the dual repumping scheme, and the peak atom number in the continuously repumped MOT is approaching to 1 billion, yielding 1$\times$10$^{11}\,$cm$^{-3}$ for the atom density. The Zeeman slower is robust as well as quick to assemble. In the $^{3}$P$_{2}$ magnetic trap, the lifetime is measured to be 1.1$\,$s. Employing resolved-sideband Zeeman spectroscopy, the Land\'{e} $g$ factor of $^{3}$D$_{1}$ is measured to be 0.4995(88) showing a good agreement with the calculated value of 0.4988, which can be further improved by locking the 2.6$\,\mu$m laser to a cavity. The results will have an impact on various applications including atom laser, dipolar interactions, quantum information and precision measurements.

\begin{acknowledgments}
We thank Alok Singh for productive discussions. We thank Richard Barron for discussing the Zeeman slower. K. B. and Y. S. acknowledge funding from the
Engineering and Physical Sciences Research Council (EPSRC) Project No. EP/T001046/1. M. M. acknowledges funding from the Defence Science and Technology Laboratory (Dstl). Q. U. acknowledges funding from the Iraqi government.
\end{acknowledgments}

\appendix
\section*{}
\nocite{*}

\bibliography{apssamp}% Produces the bibliography via BibTeX.

\end{document}